\documentclass[journal,twoside,web]{ieeecolor}
\usepackage{generic}
\usepackage{cite}
\usepackage{amsmath,amssymb,amsfonts}
\usepackage{algorithmic}
\usepackage{graphicx}
\usepackage{textcomp}
\def\BibTeX{{\rm B\kern-.05em{\sc i\kern-.025em b}\kern-.08em
    T\kern-.1667em\lower.7ex\hbox{E}\kern-.125emX}}
\begin{document}
\title{Spin-Hall MTJ Cells for Intra-Column Competition in Hierarchical Temporal Memory}
\author{Andrew W. Stephan and Steven J. Koester, \IEEEmembership{Fellow, IEEE}
\thanks{Manuscript submitted \today. This work was supported by Seagate Technology PLC.}}
\maketitle

\begin{abstract} 
We propose a dedicated winner-take-all circuit to efficiently implement the intra-column competition between cells in Hierarchical Temporal Memory which is a crucial part of the algorithm. All inputs and outputs are charge-based for compatibility with standard CMOS. The circuit incorporates memristors for competitive advantage to emulate a column with a cell in a predictive state. The circuit can also detect columns 'bursting' by passive averaging and comparison of the cell outputs. The proposed spintronic devices and circuit are thoroughly described and a series of simulations are used to predict the performance. The simulations indicate that the circuit can complete a nine-cell, nine-input competition operation in under 15 ns at a cost of about 25 pJ.
\end{abstract}


\section{Introduction}
\label{sec:introduction}
Hierarchical Temporal Memory (HTM) is an emerging neuromorphic algorithm inspired by the structural properties of the neocortex\cite{HTM1}. HTM boasts powerful recognition and prediction abilities\cite{HTMRecognition1,HTMPrediction1}. The conceptual architecture is fairly complex, with many different functions required to implement it. A comprehensive processor architecture has been proposed for this purpose\cite{Dhireesha1,Dhireesha2}. HTM consists of two primary components, the spatial pooler and the temporal memory. The spatial pooler consists of a set of columns with proximal connections to the input space. The input space is a sparse distributed representation (SDR) of data in the form of a binary matrix. Each column activates upon receiving input which exceeds its threshold value. The temporal memory portion of HTM divides each column into multiple cells that share the same proximal connections and compete with one another to represent the column. Axial connections between cells in different columns can give some cells a competitive advantage, but the proximal connections must still be solely responsible for surpassing the threshold. 

Implementing the full HTM structure in hardware is energy-expensive due to its complexity. The inclusion of dedicated circuitry capable of efficiently performing specific HTM-related tasks can reduce this load. This provides motivation to design a variable-threshold analog winner-take-all (WTA) circuit with competitive advantage. In this work we propose an efficient spintronic implementation of a WTA circuit meant to emulate the cells within an HTM column. The circuit includes an option for competitive advantage so that certain cells can be biased to win even when all of the competitors receive the same input, which emulates the 'predictive state' of HTM. We also consider how to detect a column 'bursting' which indicates that the threshold was exceeded but no cells were in a predictive state. In the following sections we will describe the device and circuit design, analyze its performance and explain our simulation methodology. 

In part the purpose of this work is to study the effect of the competitive advantage term on the operation of the spintronic WTA circuit, and determine which values should be used. This knowledge will guide future efforts, especially the choice of memristive devices needed to induce the advantage. This work does not deal with the overall HTM architecture, but focuses specifically on the proposal for an efficient implementation of individual columns. Althrough spintronic elements are involved, the input and output is voltage-based, which allows the columns to be paired with any other charge-based implementation of the other HTM functions to emulate a full HTM architecture. We use beyond-CMOS methods to develop a WTA circuit in this work to explore the potential advantages conferred by the inherently analog and non-linear nature of certain spintronic devices. As will be discussed below, we determine that the transition from CMOS to spintronic WTA implementations results in a tradeoff between delay and energy cost.

\section{Design}
\label{sec:Design}
\subsection{Spintronic Cell Design}
\begin{figure}
\centering
\includegraphics[scale=0.45]{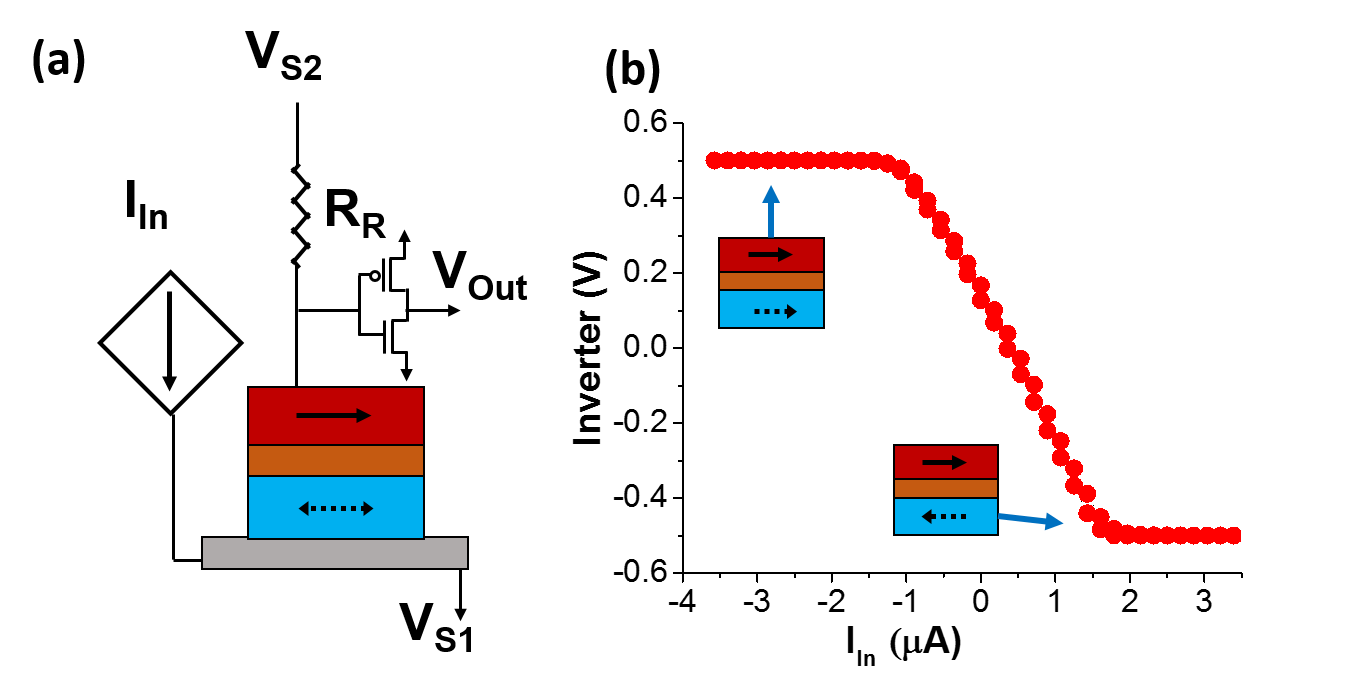}
\caption{(a) Basic cell design including MTJ, reference resistor and output inverter. (b) Output potential vs. input current in steady-state. Two different FL geometries are considered, with a step-like transition and a smooth transition.}
\label{fig:Cell}
\end{figure}

The cell is based on a well-known device, the spin-Hall-effect (SHE) driven magnetic tunnel junction (MTJ) voltage divider\cite{KRoy,mLogic,Spin Switch,Naeemi2,SHE MTJs}. The particular version of SHE-MTJ we use is derived from \cite{MAAP}, as the analog WTA circuit requires non-digital behavior from the MTJs. The MTJ free layer (FL) is in contact with a heavy metal (HM) which, when charge flows through it, produces a spin-polarized current which can be used to drive the FL. The conductance $G_{MTJ}$ of the MTJ varies with the relative angle $\phi$ of the FL magnetization to the pinned layer (PL) as
\begin{gather}
G_{MTJ} = \frac{1}{2}(G_P + G_{AP}) + \frac{1}{2}(G_P - G_{AP})cos\phi,
\label{RMTJ}
\end{gather}
where $G_P$ and $G_{AP}$ are the conductance when the FL is parallel or antiparallel to the pinned layer, respectively. The output potential of the voltage divider and ultimately that of the attached inverter\cite{MTJs1} thus depend on $\phi$. The equations governing the cell dynamics will be covered in more detail in Section \ref{sec:Simulation}. The attached inverter gives the cell a stable voltage-based read path that avoids perturbing the voltage divider (see Fig. \ref{fig:Cell}). As in \cite{MAAP}, we choose the anisotropic characteristics of the MTJ to generate a smooth linear response loop, in this case by utilizing shape anisotropy. The MTJ FL width dimension is shorter than the length dimension, creating a smooth transition due to the demagnetization field. We assign the PL orientation such that a negative current produces a spin torque in the parallel direction while positive current drives the FL in the antiparallel direction.

\subsection{WTA Circuit Design}
\begin{figure}
\includegraphics[scale=0.45]{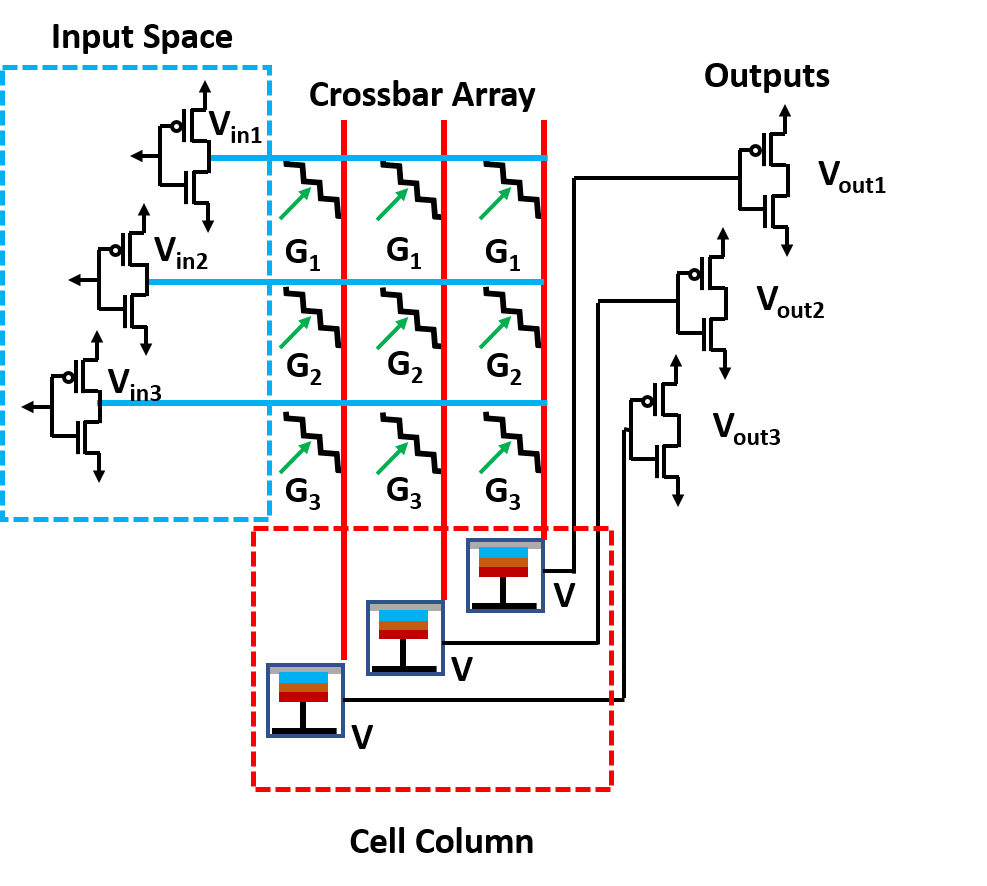}
\caption{An set of inputs connected to the column via an array of memristors. If the conductances in each row of the crossbar are identical, each cell receives the same net input as is standard in HTM.}
\label{fig:Crossbar}
\end{figure}

\begin{figure}
\centering
\includegraphics[scale=0.55]{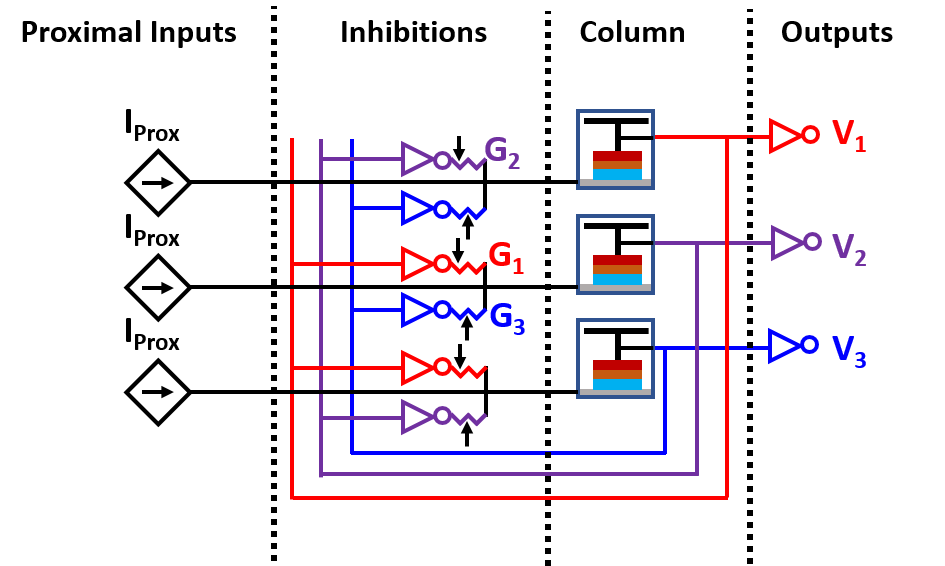}
\caption{Winner-take-all circuit design. A column consists of multiple cells, each of which receives the same proximal input current. Each cell also receives an inhibitory current from each of the other cells.}
\label{fig:WTA}
\end{figure}

\begin{figure}
\includegraphics[scale=0.45]{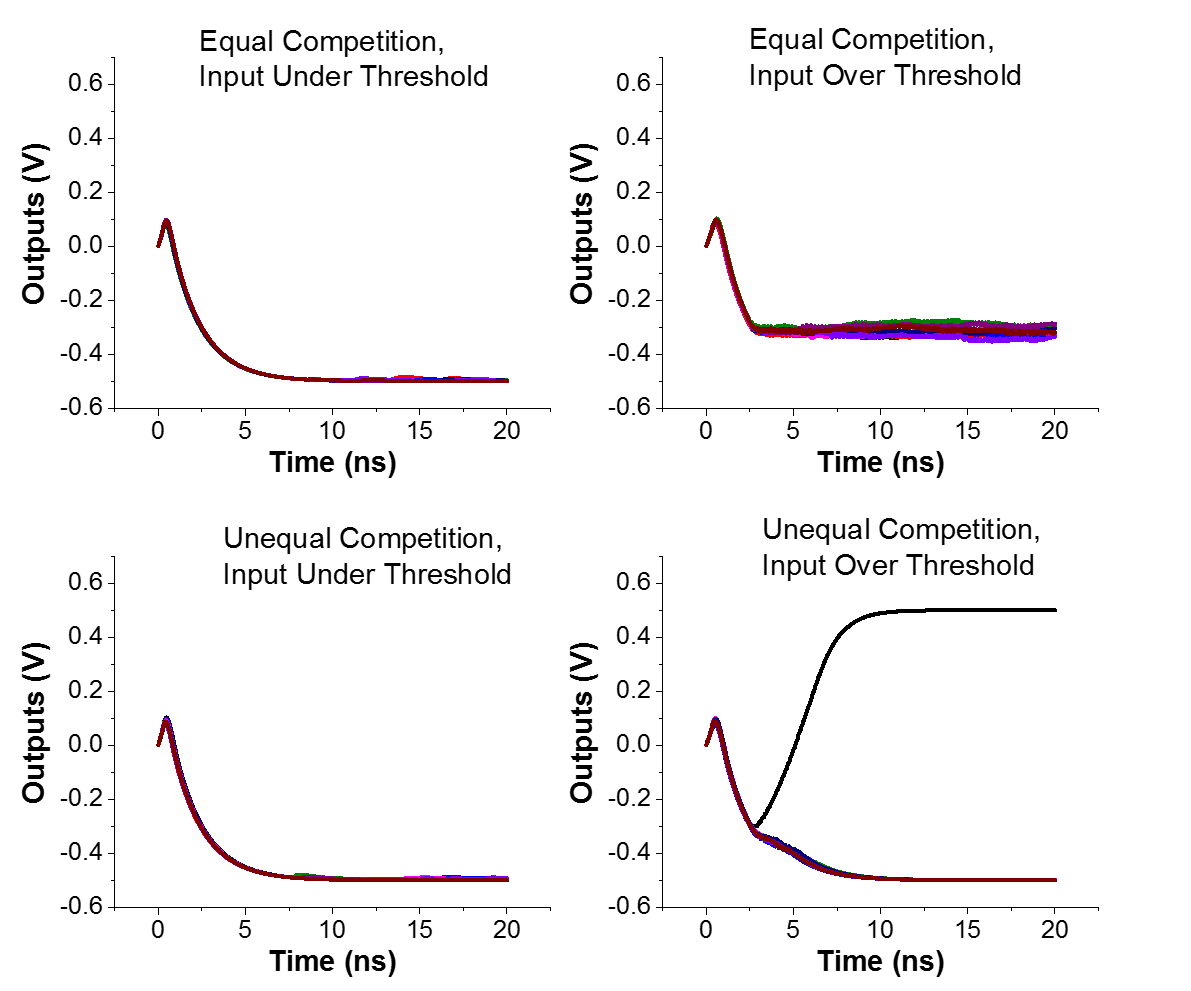}
\caption{Outcomes for four different basic cases. Each case assumes a 9-cell column competing with identical inputs. When the input is insufficient to excite the cells, all cells quickly reach a -0.5 V output and remain there. When the input is sufficient and the cells compete on equal terms, all cells achieve an equilibrium with one another at an above-minimum output. When the input is sufficient and the cells compete on unequal terms, the cell with advantage quickly drives the others to the minimum output.}
\label{fig:Competition}
\end{figure}

An HTM column consists of multiple cells that receive the same proximal input and compete in a WTA fashion to represent the column if the input exceeds some threshold. The threshold may be tuned by an input bias. In this work we draw much of the WTA cell design from \cite{MAAP}. The workings of individual cells is studied in detail in that work. For this application, we simplify the pooler circuit by removing the second stage of each activation pair because the neural activation function is not needed. We assume the input space takes the form of a set of voltage sources. Current is provided to the cells by a memristor crossbar array joining the cells with the input space. An example is shown in Fig. \ref{fig:Crossbar}. The precise nature of the memristors is not treated here but many examples exist in the literature including filamentary, MTJ-based and ferroelectric memristors\cite{Filamentary1,Filamentary2,MTJ1,FE1}, any of which would be suitable for this purpose. Some architectures also incorporate an intermediary device which reads the input and transmits a corresponding signal to the cells. Besides the current from the proximal connections, each cell has an additional connection from each of the output inverters of its neighbors. The result is an inhibitory connection between each cell that induces more negative torque in the receiver cell as the magnetization of the source cell becomes more positive. The strength of the inhibitory connections depends on the conductance of the memristor joining each output inverter to the HM input. A higher conductance gives the source cell a competitive advantage, which is measured as the ratio of the conductance to that of the other cells. The complete WTA circuit design is shown in Fig. \ref{fig:WTA}, where the crossbars are represented as simple current sources and the inverters are represented by the standard circuit symbol for brevity. The operating parameters are carefully chosen such that the low sensing potential $V_{S1}$ on the voltage divider and the inverter low rail potential $-V_{DD}$ are matched. The result is that the inhibitory output connections cease to provide current when the source cell magnetization reaches the $-1$ state. This allows the cells to achieve an equilibrium by balancing the excitatory proximal connections with the intra-column inhibitory connections. Example results for four different cases are given in Fig. \ref{fig:Competition}. If all cells compete equally when receiving a negative input current sufficiently large to excite them, they all achieve a similar steady-state output below the value expected according to the proximal input but above the minimum output. Alternatively if the competition is uneven due to a certain cell being in a predictive state, that cell drives the others to the minimum state while itself achieving a higher state. This result is achieved by giving the predictive cell a higher output conductance on its inhibitory connections. A detailed breakdown of the WTA circuit performance is given in Section. \ref{sec:Results}.

\section{Results}
\label{sec:Results}
There are two crucial questions that determine the success of this WTA circuit in its intended function for the HTM architecture. The first is the question of whether it can emulate a predictive state via competitive advantage for one cell. The second is the question of whether it can emulate a column bursting. This situation occurs in an HTM when the proximal input is large but none of the cells is in a predictive state. The discussion below resulted from a series of simulations of the full 9-cell circuit using a custom simulator written in Matlab, which includes empirical approximations of the inverter behavior based on previous HSPICE simulations\cite{MAAP}. 

\subsection{Predictive State}
\begin{figure}
\includegraphics[scale=.45]{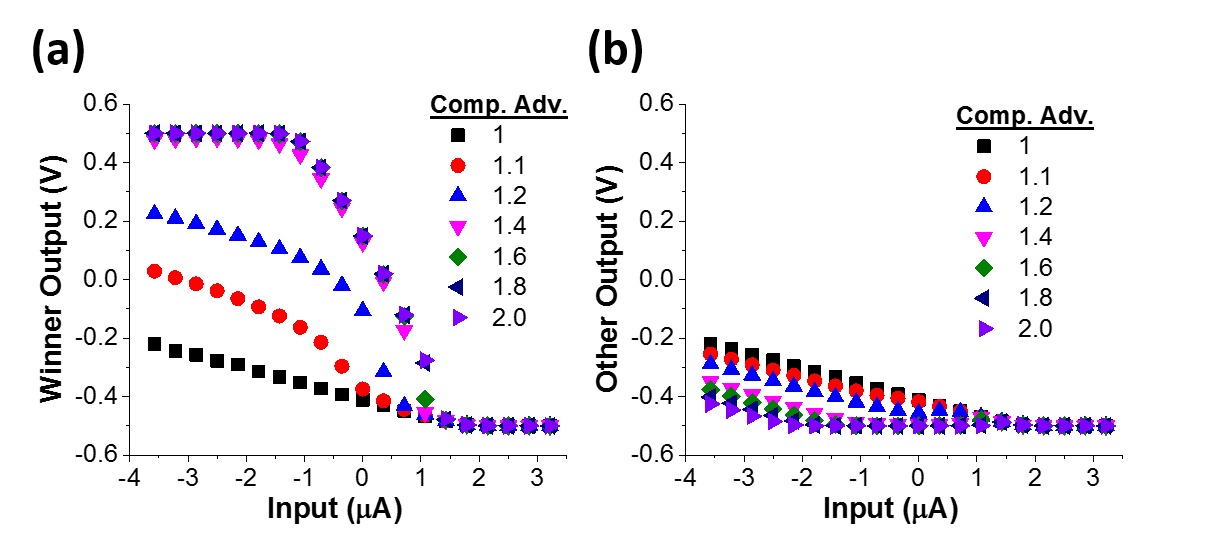}
\caption{(a) Predictive cell output vs. input current. (b) Average of other cell outputs vs. input current.}
\label{fig:PredOut}
\end{figure}

\begin{figure}
\includegraphics[scale=.7]{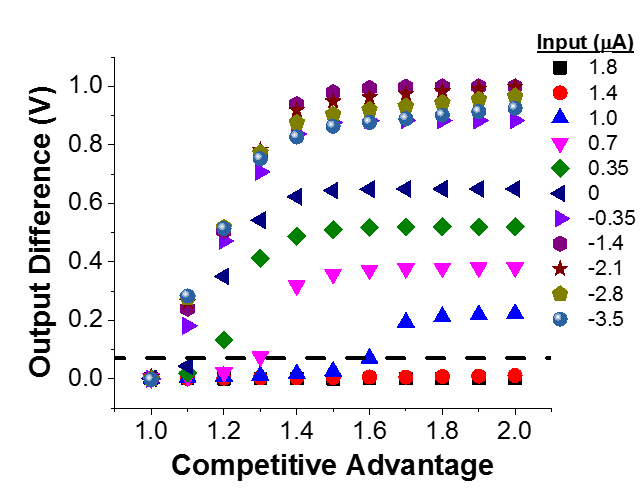}
\caption{Difference between the predictive cell output and the average of the other cell outputs for various inputs as a function of competitive advantage. The dashed line indicates the minimum required separation of 70 mV.}
\label{fig:PredDiff}
\end{figure}

\begin{figure}
\includegraphics[scale=0.7]{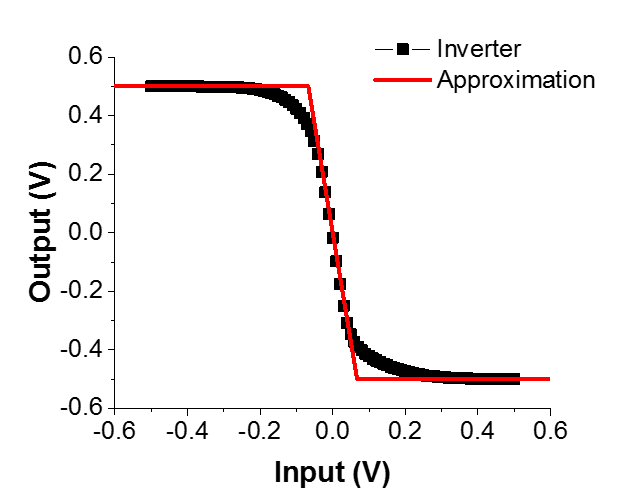}
\caption{Inverter behavior based on HSPICE simulation with linear approximation included.}
\label{fig:Inverter}
\end{figure}

To determine whether the WTA circuit can succeed in emulating a predictive state, we performed a series of Monte-Carlo simulations and averaged the results. Each 100-round ensemble assumed a specific proximal input value and a competitive advantage, encoded as the ratio between the inhibitory output conductance of the predictive cell and that of the other cells. In Fig. \ref{fig:PredOut}(a) we show the designated predictive cell output vs. the input current. The larger the competitive advantage is, the greater the winner output grows as a function of input magnitude. Meanwhile in Fig. \ref{fig:PredOut}(b) we show that as the competitive advantage grows, the other cell outputs shrink. Fig. \ref{fig:PredDiff} shows directly the difference between the winner cell and the mean of the others, as the other cells deviate very little from one another. The dashed line indicates a difference of 70 mV, which is the smallest signal difference which is still large enough for a digital inverter to differentiate. With a reference potential of $V_{S1} + 35$ mV, the inverter in Fig. \ref{fig:Inverter} can produce outputs that differ by 500 mV using inputs of $V_{S1}$ and $V_{S1} + 70$ mV. Here we note that the choice of competitive advantage can be used to determine the effective input threshold for detection of a predictive state event. In order to noticeably differentiate even at weak excitation inputs such as $+1 \mu A$ (see Fig. \ref{fig:Cell}),  a competitive advantage of at least 1.6 is required. This yields sufficient output separation to differentiate the winner cell from the others. Alternatively, a low advantage of 1.1 can be chosen in order to enforce a 0 $\mu A$ threshold because at this advantage level only negative currents are shown to produce results which exceed $V_{S1} + 70$ mV. In general the current threshold which is enforced depends on the magnitude of the competitive advantage used. 

\subsection{Bursting}
\begin{figure}
\includegraphics[scale=.7]{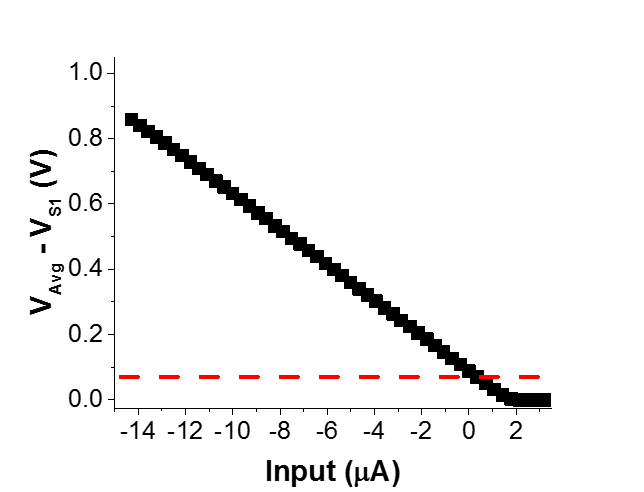}
\caption{Difference between the column average output the minimum cell potential vs. input current. The dashed line indicates the minimum required separation of 70 mV.}
\label{fig:Bust}
\end{figure}

To simulate a column going bust, we assume no cells are in the predictive state, which corresponds to a competitive advantage of 1. As shown in Fig. \ref{fig:Competition}, the cells will all behave identically in the case of going bust, reaching a steady-state slightly above the minimum output. This can be detected by measuring at least two outputs. We assume the mean output of all cells is measured, and the column is determined to be bust if the average exceeds some threshold but no single cell has a large output. To estimate the detection capability, we assume that this average output must exceed $V_{S1}$ by at least 70 mV, sufficient for a digital inverter to differentiate the signals as mentioned above. Fig. \ref{fig:Bust} shows the average potential difference vs. input. We note that if the input is below 0 $\mu A$, which would suffice to excite the cells if not for the inhibitory connections, a passive averaging circuit can detect that the column has gone bust. We note that while $V_{Avg} - V_{S1}$ can be expected to exceed 70 mV in cases with a single predictive cell, a simple digital comparison circuit can differentiate those cases from the ones in which all cells are partially excited. 

\subsection{Energy Usage}
\begin{figure}
\includegraphics[scale=.7]{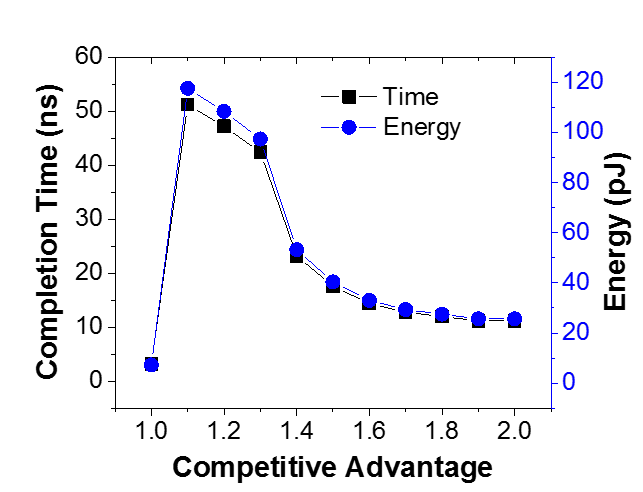}
\caption{Time and energy required to complete the WTA function vs competitive advantage. Greater advantage leads to a faster solution.}
\label{fig:Energy}
\end{figure}

The average time to complete a WTA function depends on the competitive advantage of the predictive cell. If there is no predictive cell--or no competitive advantage--then the process is quite fast, requiring about 3 ns. At most the WTA process takes less than 60 ns to finish. If there is a predictive cell, then the process finishes more quickly if the advantage is larger, as shown in Fig. \ref{fig:Energy}. This is because the additional advantage causes the predictive cell to drive its competitors with more current, suppressing them more swiftly. When there is no competitive advantage, the process finishes most quickly since the cells quickly reach an equilibrium. In this case there is no need to wait for one cell to differentiate itself from the rest by suppressing their outputs, as all cells behave nearly as one, barring noise. The energy consumption is also given in that figure. The cost is at most 120 pJ for a nine-cell column, or about 13 pJ per cell. The relevant power calculations are given in Section \ref{sec:Simulation} C.

\subsection{Process Variation}
\begin{figure}
\includegraphics[scale=.7]{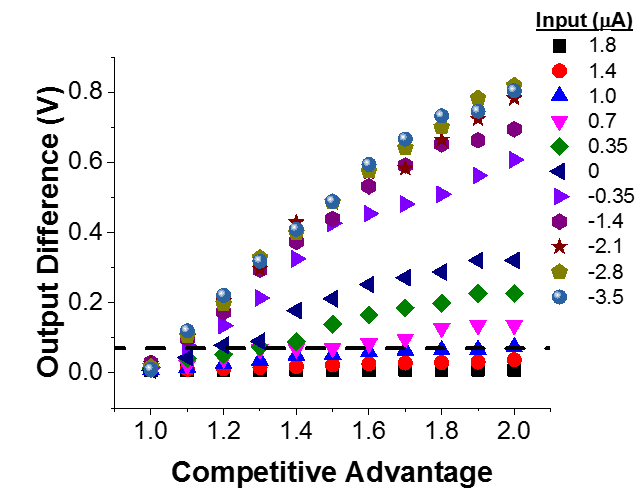}
\caption{Difference between the predictive cell output and the average of the other cell outputs for various inputs as a function of competitive advantage. The dashed line indicates the minimum required separation of 70 mV. This simulation accounts for process variation in the transistors and MTJs of the WTA circuit.}
\label{fig:PredDiffVar}
\end{figure}

Here we consider the effects of process variation on the outcome of the WTA cell in the predictive state case using Monte-Carlo simulations. Each device in the simulated circuit is randomly assigned a set of parameters drawn from normal distributions at the beginning of every round of simulation. As before, each data point is the mean result of an ensemble of at least 100 rounds. We consider four different independent variables: transistor threshold, MTJ parallel resistance, MTJ antiparallel resistance and MTJ base switching current $I_{c0}$, which is the value of $I$ at which $\boldsymbol{H_{SHE}}$ equals $\boldsymbol{H_I}$ (\ref{HEFF}-\ref{Ispin}). As in \cite{MAAP}, Typical standard deviations of 5\% for each MTJ parameter were chosen after consulting \cite{Variation1,Variation2} and a transistor threshold deviation of 20 mV was selected based on the Pelgrom plots in \cite{Variation3}, assuming 200 nm gate width. Fig. \ref{fig:PredDiffVar} shows the output difference between the winner cell and the others with competitive advantage as a parameter as in Fig. \ref{fig:PredDiff}. Incorporating process variation makes it somewhat more difficult for the predictive cell to differentiate itself from the others. A competitive advantage of 2.0 is required when a weak proximal excitation of 1$\mu A$ is applied, compared to 1.6 when ideal devices are used. Similarly an advantage of 1.2 is required to enforce the 0 $\mu A$ threshold as opposed to 1.1 with ideal devices. While these differences are notable, they indicate that the circuit is not prevented from proper function by process variation so long as the modified advantage requirements are applied. We also note that simulation of a column going bust with process variation incorporated showed no noticeable difference in results.

\section{Simulation Methods}
\label{sec:Simulation}
The physical parameters used in the simulation are available in Table \ref{tab:Parameters}. In choosing the magnetic saturation, weak in-plane crystalline anisotropy, MTJ resistance and TMR, we selected values typical for spintronic devices after consulting \cite{SHE MTJs,MTJs1}.

\begin{table}
\centering
\caption{Simulation Parameters}
\label{table}
\setlength{\tabcolsep}{3pt}
\begin{tabular}{|p{25pt}|p{100pt}|p{75pt}|}
\hline
Symbol& 
Quantity& 
Value \\
\hline
\vspace{0.005in}
$K$& 
\vspace{0.005in}
crystalline anisotropy& 
\vspace{0.005in}
10 kJ/m$^3$ \\
$V$& 
ferromagnet volume& 
1800 nm$^3$ \\
$M_S$& 
saturation magnetization& 
1 MA/m \\
$\alpha$&
Gilbert damping&
0.01 \\
$t_{HM}$ &
heavy metal thickness &
5 nm \\
$R_{HM}$ &
heavy metal resistance &
$\approx$ 50 $\Omega$ \\
$\theta$ &
spin-Hall angle &
0.3\cite{Tungsten,Tantalum} \\
$V_{S1}$ &
low sensing voltage&
-0.5 V \\
$V_{S2}$ &
high sensing voltage&
0.4 V \\
$R_M$ &
MTJ RA product &
8 $\Omega$ $\mu m^2$ \cite{MTJs1}\\
$TMR$ &
tunnel magnetoresistance ratio &
1.5 \cite{MTJs1}\\
$R_R$ &
reference resistor &
3.25 - 140 k$\Omega$ \\
$\Delta$t &
simulation time step &
0.5 ps \\
$V_{DD}$ &
inverter rail voltage &
$\pm$ 0.5 V \\
$\tau$ &
inverter intrinsic delay &
4.5 ps \\
$C_g$ &
inverter gate capacitance &
6.6 fF \\
$R_T$ &
inverter on-resistance &
10.8 k$\Omega$ \\
\hline
\multicolumn{3}{p{200pt}}{}\\
\end{tabular}
\label{tab:Parameters}
\end{table}

\subsection{MTJs}
The cell is simulated using the fourth-order Runge-Kutta method to predict the behavior of the circuit and magnetic FL. The FL is treated using the macrospin approximation and the Landau-Lifshitz-Gilbert (LLG) equation
\begin{gather}
\frac{d\boldsymbol{\hat{m}}}{dt} = -\gamma \mu_0 \Big( (\boldsymbol{\hat{m}} \times \boldsymbol{H_{Eff}}) - \alpha \big(\boldsymbol{\hat{m}} \times (\boldsymbol{\hat{m}} \times \boldsymbol{H_{Eff}})\big)\Big),
\label{LLG}
\end{gather}
where $\boldsymbol{\hat{m}}$ indicates the unit magnetization of the FL and $H_{Eff}$ is the effective field on the FL. The symbols $\gamma,$ $\mu_0$ and $\alpha$ are the gyromagnetic ratio, vaccuum permeability and Gilbert damping respectively. Bold font indicates vector quantities. The effective field consists of two terms,
\begin{gather}
\boldsymbol{H_{Eff}} = \boldsymbol{H_I} + \boldsymbol{H_{SHE}},
\label{HEFF}
\end{gather}
where $\boldsymbol{H_{SHE}}$ is the effective SHE-field and $\boldsymbol{H_I}$ represents all the intrinsic field terms. $\boldsymbol{H_{SHE}}$ is proportional to the spin current $I_S$\cite{Ralph and Stiles,Ispin}:
\begin{gather}
\boldsymbol{H_{SHE}} = \frac{1}{\mu_0}\frac{I_S}{2q}\frac{\hbar}{\alpha V M_S} \boldsymbol{\hat{y}},
\label{HSHE}
\end{gather} 
where $\hbar$ is the reduced Planck constant. The spin current $I_S$ is in turn proportional to the charge current $I$ flowing through the HM layer:
\begin{gather}
I_S = \theta \frac{t_{HM}}{L_{FM}} I,
\label{Ispin}
\end{gather}
where $\theta$ is the spin-Hall angle. The intrinsic field $\boldsymbol{H_I}$ includes the demagnetization and anisotropy fields as well as the thermal noise field which consists of a multivariate Gaussian random variable with zero mean and variance
\begin{gather}
\sigma_T = \sqrt{\frac{2k_BT\alpha}{\gamma M_SV\Delta t}},
\label{HT}
\end{gather}
where $k_B$, $T$, $M_S$, $V$ and $\Delta t$ are the Boltzmann constant, temperature, magnetic saturation, FL volume and simulation time step respectively. We account for this term as it can cause noise in the cell voltage readouts.

\subsection{Inverters}
The voltage dividers provide the gate potential for the inverters and are treated with standard circuit equations while taking into account the changing $RC$ gate delay due to the varying MTJ resistance. The output of the inverter is treated with a first-order approximation based on HSPICE simulations using the 16-nm node Predictive Technology Model\cite{PTM} (see Fig. \ref{fig:Inverter}). The inverter gate width for the HSPICE simulations was 200 nm. 

\subsection{Energy calculations}
The power consumed by the WTA circuit comes from three sources: inverter rail-to-rail leakage $P_I$, crossbar input power $P_{CB}$ and voltage divider leakage $P_{VD}$. Assuming nine cells in a column with nine powered connections to the input space each, there are nine voltage divider stacks and 162 inverters, so $N_{VD} = 9$ and $N_{I} = 162$. Using HSPICE simulations to estimate the rail-to-rail leakage we find that $P_I = 7.3 \mu W$. Assuming an average inhibition output resistance of $R = 280$ $k\Omega$, each crossbar connection drains an average of $P_{CB} = \frac{(E[V_{In} - V_{S1})^2]}{R} = 1.2 \mu W$. Finally, the voltage divider leakage is estimated as $P_{VD} = \frac{(V_{S2} - V_{S1})^2}{R_R + E[R_{MTJ}]} = 30.4 \mu W$. The overall WTA delay $\tau$ is estimated as the time at which the average output comes to within 5 mV of its steady-state value. The total energy cost is $E = \tau \cdot \big(N_I\cdot(P_I + P_{CB}) + N_{VD}\cdot P_{VD}\big)$. The distribution of $\tau$ and $E$ is shown in Fig. \ref{fig:Energy}. We note of course that the values in Fig. \ref{fig:Energy} vary depending on the number of cells per column and the average number of connections each column has to the input space.

This work is comparable to \cite{WTA1, WTA2, WTA3} which describe purely CMOS-based WTA implementations. Although the circuit in this work consumes more energy per operation, it requires significantly less time per input set. We ascribe this difference to the several additional layers of computation required by the CMOS circuits which introduce more delay. However, spintronic WTA circuit involves more leakage current, which accounts for the increased energy cost despite its reduced runtime.

\section{Conclusion}
\label{sec:Conclusion}

The HTM algorithm is a powerful recognition and prediction tool with the potential to revolutionize neuromorphic systems. Each portion of the HTM algorithm that can be implemented using efficient dedicated circuits significantly reduces the overall computational burden. The intra-column dynamics are an important part of HTM, and we have demonstrated a novel spintronic circuit based upon spin-Hall MTJs with a simple design that can emulate these dynamics quickly and efficiently. To the best of the authors' knowledge, there are no proposals for column circuits which more efficiently implement the intra-column competition aspect of HTM.

\end{document}